\documentstyle[palatino,psfig,11pt]{article}
\begin{document}

\begin{center}
{\Large Effects of core degrees of freedom in \\
the low energy $^7$Be$(p,\gamma)^8$B reaction}\\
\vspace{0.2cm}
F.M. Nunes$^{1}$, R. Crespo$^1$, I.J. Thompson$^2$ \\
\vspace{0.2cm}
{$^{1}$ Departamento de F\'{\i}sica, CENTRA, IST, Lisboa, Portugal.}  \\
{$^{2}$ Department of Physics, University of Surrey, GU2 5XH, U.K.} 
\end{center}

\begin{abstract}
Quadrupole and octupole couplings to all bound states
of $^7$Be are included  in describing the capture reaction $^7$Be$(p,\gamma)^8$B.
We verify, contrary to what we had initially stated, that the energy behaviour of the Astrophysical S-factor in the energy 
range of interest is not significantly sensitive to the core couplings in the 
g.s. (g.s.) of $^8$B
although its overall magnitude is shifted. We find there is some sensitivity to 
the various deformation models when introducing the nuclear interaction
for calculating the scattering states.
The various deformation models predict quite different contributions to the quadrupole
and magnetic moments but in order to compare with the data the
quadrupole and magnetic moment of $^7$Be need to be measured.

\end{abstract}

The importance of reliable extrapolations of the proton capture reactions 
from the measurable  energy range to astrophysical energies is
so far unquestionable although experiments have been 
reaching lower and lower in energy.
Up to the present year it was fairly well established
that the low energy behaviour
of the Astrophysical S-factor
was determined by the asymptotic of the g.s. wavefunction and therefore
was not dependent on any particular
nuclear model \cite{xu}.
In this letter  we reformulate our results of \cite{nunes1} which indicated  that core quadrupole couplings 
in the g.s. of
$^8$B could introduce
modifications in the low energy behaviour of the  Astrophysical S-factor.
Due to a numerical error this was found not to be correct.
 We present extended results of
calculations including all bound states of the core ($^7$Be) and test
the effect of  octupole as well as quadrupole couplings.
In addition we clarify the role of the scattering nuclear interactions
in the evaluation of the S-factors.

For a fair evaluation of the real effect of core deformation,
we chose the same radius and
diffuseness as Tombrello \cite{tom}, Robertson \cite{rob} and Kim \cite{kim}. We use the same 
spin orbit depth as Kim. The depth of the central interaction  is adjusted
to reproduce the correct binding energy.
The deformation of the $^{7}$Be core is taken to be $|\beta_2|=0.5$
which corresponds to a quadrupole moment  slightly larger than the 
value for $^{7}$Li ($Q=-4.06 \pm 0.08$ e fm$^2$ \cite{ajen}). 
According to microscopic calculations
\cite{baye,csoto} the deformation parameter of $^{7}$Be could actually be larger ($\beta_2=0.6-0.7$), enhancing the effects.
Since there is no experimental indication whether $^7$Be is prolate or oblate
we have considered both possibilities.  
We have performed the coupled channel calculations for p+$^7$Be,  
 for the case of reorientation only,  and when  both the $\frac{1}{2}^-$ state and the $\frac{7}{2}^-$ state of $^7$Be are included. 
Due to the $^4$He$-^3$He cluster nature of $^7$Be one would expect
the octupole deformation of the core to be non zero. We have also looked into this possibility. 
In table \ref{pardeftb} we present a range of possible models to describe this system.
The probability of finding the system in some particular channels is presented in that
same table.
The admixtures become very large when core excited  states are included.
Inclusion of octupole couplings produces a weak effect,
although still reducing the admixture caused by the quadrupole deformation.
From all cases we find that both the density distribution and the momentum
distribution do not differ significantly from those of the inert core models \cite{nunes1}
and the results are summarised in table \ref{deftb}.
Core deformation tends to increase the radius which reflects the effective 
increase of volume of the interaction. We verify that in all cases
the rms matter radii predicted are slightly too large \cite{tanrms,jim}.
The widths are about twice the FWHM obtained from experiment \cite{schwab}
underlining the need for  a proper breakup reaction theory \cite{obuti96}.
We verify that each channel of the wavefunction reaches its asymptotic
limit for $r \leq 7$ fm to $0.1\%$.

The quadrupole moment of $^8$B has been measured \cite{minamisono}
($|Q|=6.83 \pm 0.21$ e fm$^2$).
but the quadrupole moment of $^7$Be is not known. 
In our model we can calculate the quadrupole contribution 
$Q_R$ due to the valence proton.
In the single channel model, the coupling of the angular momentum implies $Q_R=0$.
Due to the algebra, it is the overlap between $[p_{3/2} \otimes \frac{3}{2}^-]$ and
$[p_{1/2} \otimes \frac{3}{2}^-]$ channels that basically determines
the strength of $Q_R$.
In table \ref{qmtb} we present the contributions from the valence proton of $^8$B 
to its quadrupole moment within our set of deformed core models.
The final prediction for the $^8$B
quadrupole moment relies on the core's 
quadrupole moment and B(E2) between the core's bound states, as yet unmeasured.
The core  contribution is mainly proportional to the probability  
of the $[p_{1/2} \otimes \frac{3}{2}^-]$ channel. 
As an example and in order to quantify the core's contribution,
if one takes $Q(^7$Be) as predicted  by GCM calculations \cite{baye}
one would need $\beta_2 \simeq 0.25$ to  reproduce $|Q(^8$B$)|$
within a reorientation model.

In the same way, the magnetic moment of $^8$B is established \cite{ajen} 
($\mu = 1.0355 \pm 0.0003 \mu_B$) but the magnetic moment of $^7$Be 
has not been measured.
In table \ref{qmtb} we present the contributions of the 
valence proton to the total magnetic moment of the g.s. of $^8$B.
Given the sensitivity of the models to these observables, the measurement of
 the quadrupole and magnetic moments of $^7$Be is very important.

\begin{figure}[h]
\centerline{
	\parbox[t]{0.7\textwidth}{
\centerline{\psfig{figure=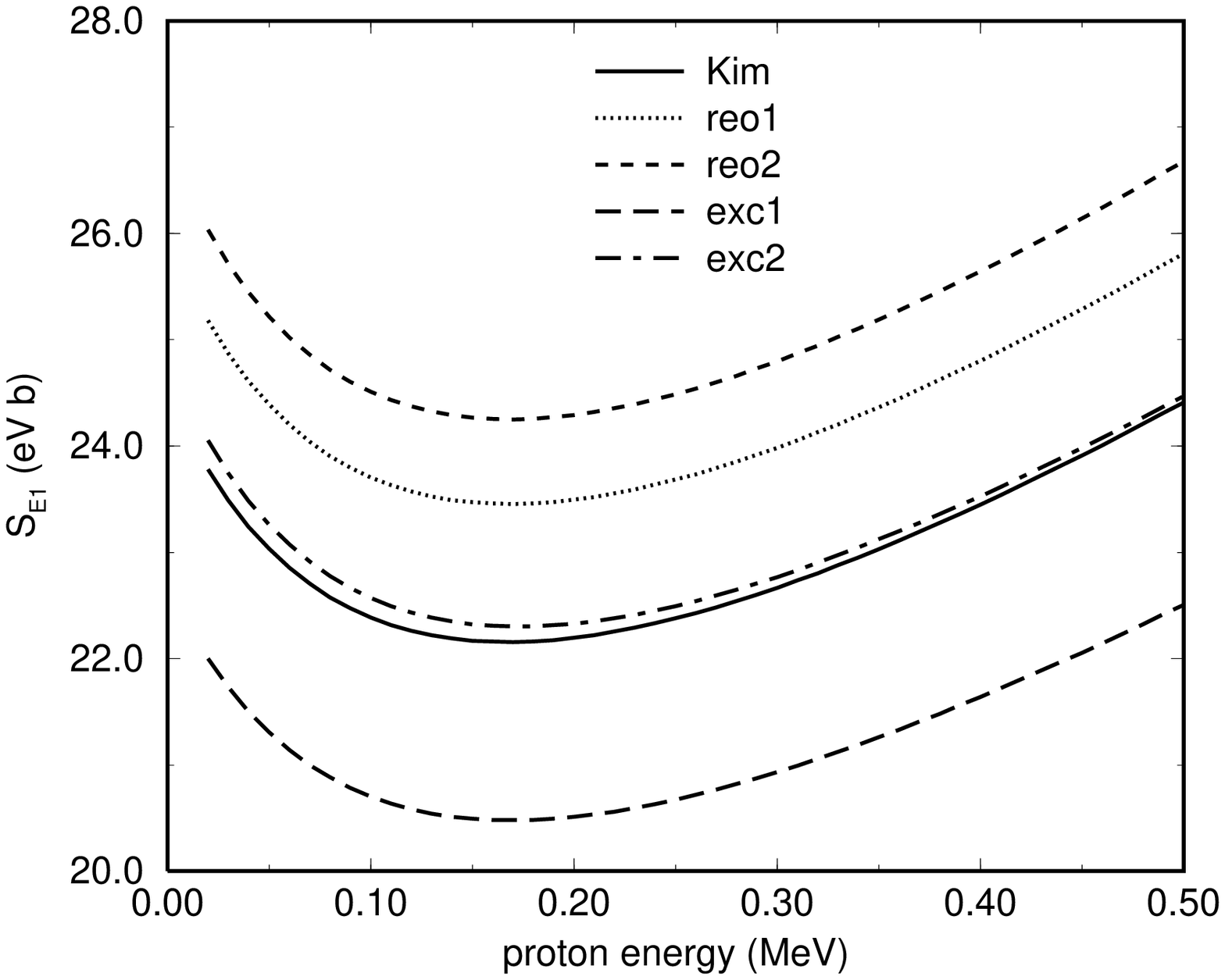,width=0.7\textwidth}}
	\caption{The variation of the S(E=20 keV) with the quadrupole
deformation parameter. }	
\label{b8e1all}}
}
\end{figure}

Reorientation couplings increase the value for the S-factor due to the 
effective volume increase (radius and diffuseness) caused by deformation (fig.\ref{b8e1all}). 
When core excitation is introduced, the overall normalisation of the
wavefunction is taken from the main g.s. channel into other components
that have a more rapid radial decay.  One would then expect a reduction of
the dipole distribution, and a decrease of the S-factor.

In table \ref{e1tb} we present the $S_{E1}$ for $E=20$ keV and $E=100$ KeV, followed by
the ratio $\frac{S(E=20)}{S(E=100)}$ in order to give insight to the energy behaviour. 
As can be immediately seen all models produce the same
energy behaviour for this energy range. Contrary to what we stated in the 
original paper \cite{nunes1}, we conclude that the structure of the core does not alter
in a significant way the shape of the energy behaviour
of the S-factor.
Microscopic models appear
to corroborate the same energy dependence \cite{kolbe}. 

\begin{figure}[h]
\centerline{
	\parbox[t]{0.7\textwidth}{
\centerline{\psfig{figure=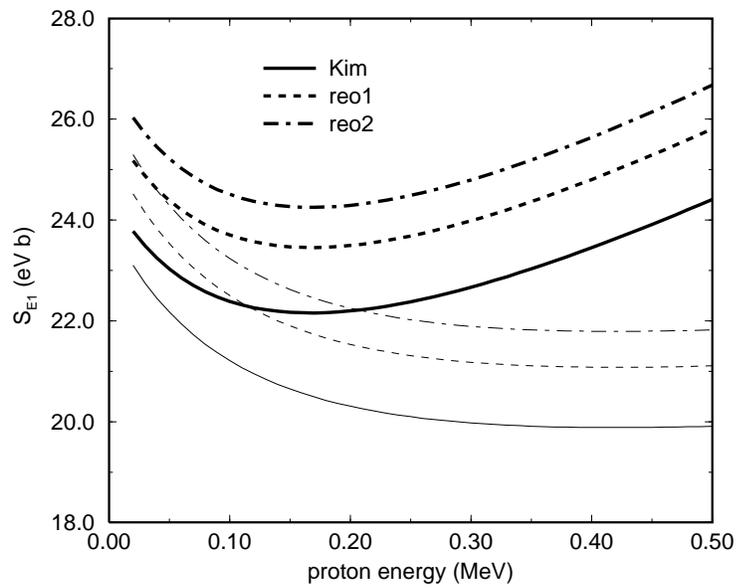,width=0.7\textwidth}}
	\caption{The effect of neglecting the contribution of the
interior to the dipole S-factor when
neglecting the nuclear interaction in the continuum:
the thick lines correspond to the full calculation whereas the thin lines
correspond to a radial cutoff of $7$ fm. }
\label{b8e1anc}}
}
\end{figure}

Given that the g.s. wavefunction reaches its
asymptotic form for $r > 7$ fm we have repeated these calculations introducing
a radial cutoff at $r=7$ fm and comparing with the full calculations.
The results are shown in fig.(\ref{b8e1anc}).
As expected, we find that the contributions from the interior are not important
(of the order of $4\%$) for $E \simeq 20 $ keV in agreement with \cite{xu}.
The interior contribution increases rapidly with increasing proton energy.

\begin{figure}[h]
\centerline{
	\parbox[t]{0.7\textwidth}{
\centerline{\psfig{figure=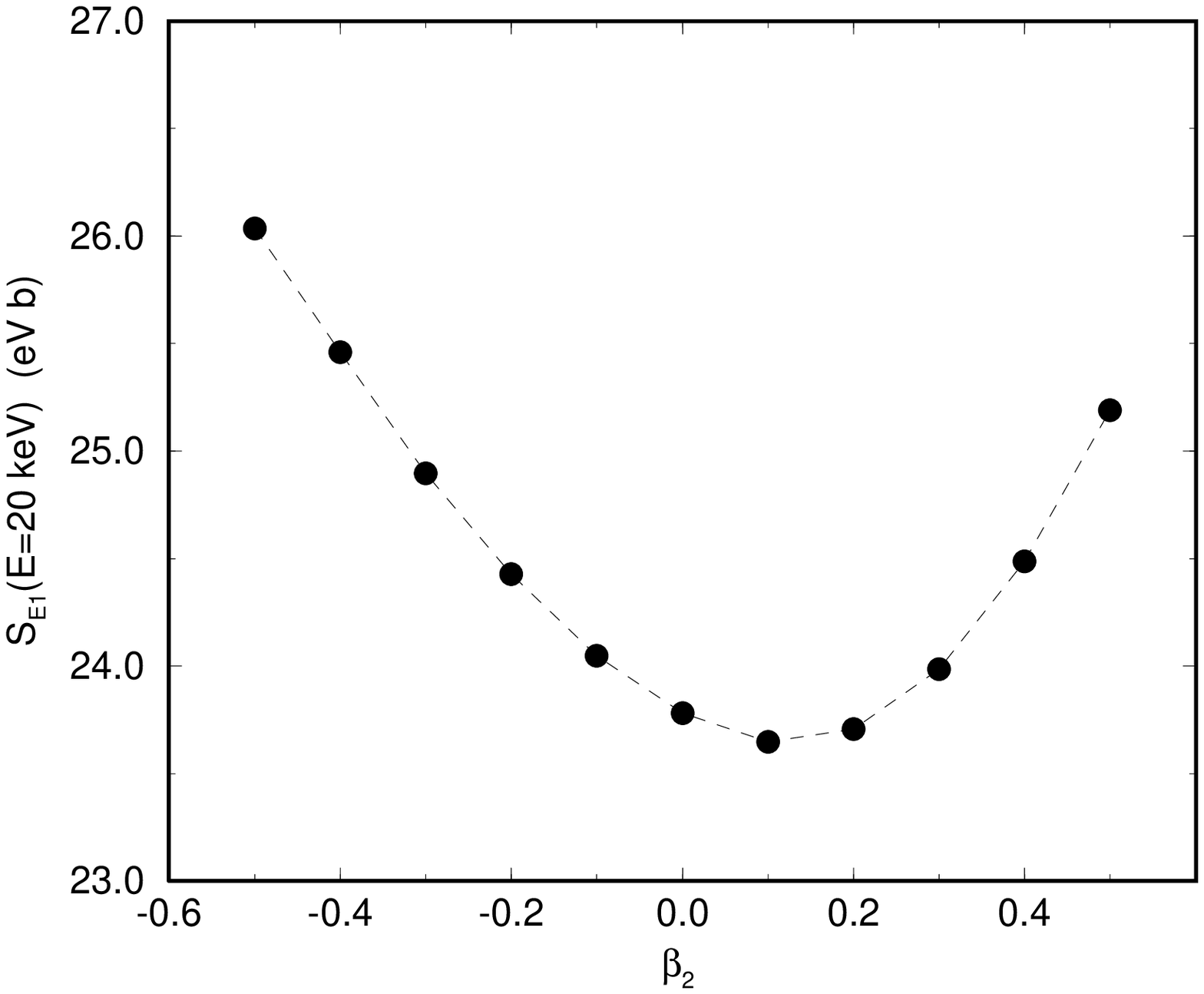,width=0.7\textwidth}}
	\caption{The variation of the S(E=20 keV) with the quadrupole
deformation parameter. }
\label{b8beta}}
}
\end{figure}

A plot of the S-factor at $E=20$ keV as a function of the 
deformation parameter is presented in fig.(\ref{b8beta}).
The calculations include only E1 transitions and neglect the nuclear interaction
in the scattering calculations.
There is a non-linear (nearly quadratic) dependence of S($20$ keV) 
with $\beta_2$ due to the volume increase of the core nucleus. 
In opposition to this effect there is the reduction of normalisation associated
with the main contributing channel $[p_{3/2} \otimes \frac{3}{2}^-]$.
It is not clear how this behaviour relates to the result obtained in \cite{csoto}
where the S-factor is found to have a linear dependence on the
quadrupole moment.

\begin{figure}[h]
\centerline{
	\parbox[t]{0.7\textwidth}{
\centerline{\psfig{figure=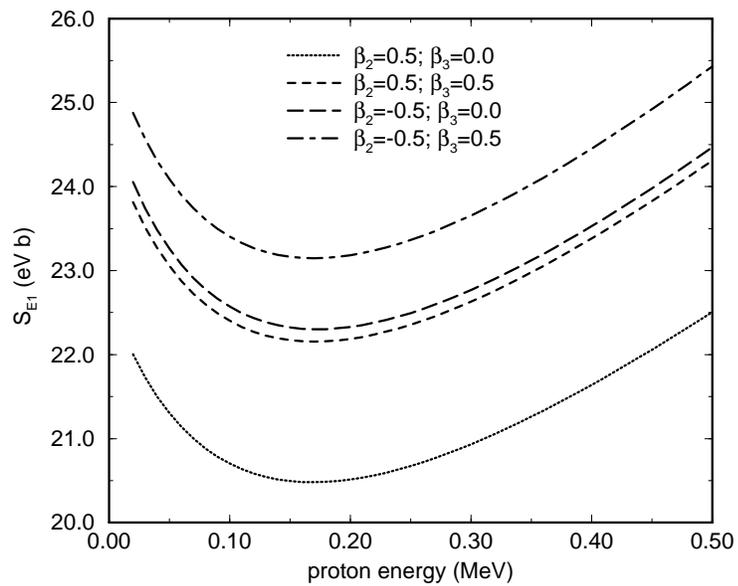,width=0.7\textwidth}}
	\caption{The effect of the octupole couplings on the
low energy S-factor.}
\label{b8octo}}
}
\end{figure}

In fig.(\ref{b8octo}) we show the effect of introducing octupole
deformation in the description of the core (as an example we take $\beta_3=0.5$).
Note that because all core states have the same parity
$\langle \phi_i |  \hat O_3 | \phi_j \rangle = 0$,
so for this system  octupole deformation produces its effects
through the quadrupole and hexadecapole form factors.
The introduction of the $\beta_3\neq0$ produces an increase in the overall normalisation of the S-factor, which again is partially explained 
by an increase in the volume of
the core ($R_{ws}$ and $a_{ws}$ have been kept equal to Kim's parameters)
but is also caused by the reduction of the excited core components of the g.s.
wavefunction.

\begin{figure}[h]
\centerline{
	\parbox[t]{0.7\textwidth}{
\centerline{\psfig{figure=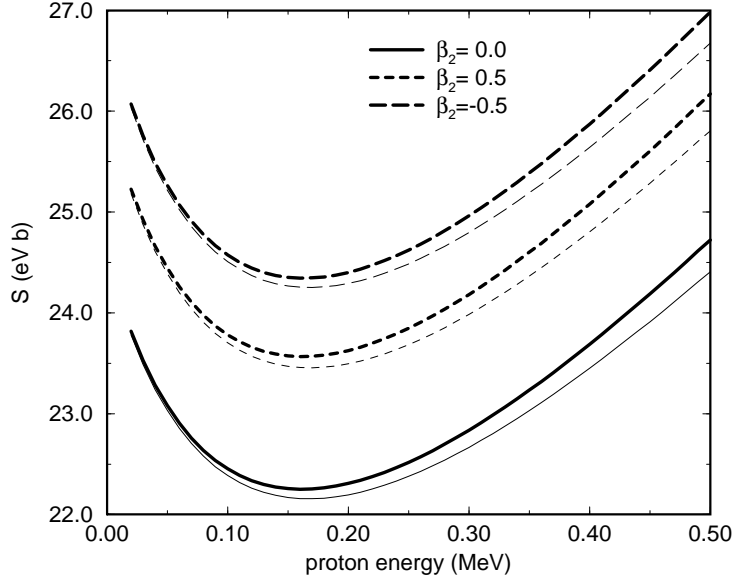,width=0.7\textwidth}}
	\caption{The comparison of $S_{tot}(E)$ including
E1, E2 and M1 transitions (thick lines) and $S_{E1}(E)$ only due to E1 capture
(thin lines).}	
\label{b8e1e2m1}}
}
\end{figure}

So far we have always neglected E2 and M1 effects. In fig.(\ref{b8e1e2m1})
we compare calculations performed for the S-factor
including only E1 transitions, $S_{E1}$,
versus those including E2 and M1 contributions as well, $S_{tot}=S_{E1}+S_{E2}+S_{M1}$
(not taking into account the nuclear interaction for the scattering states).
Although at very low energy there is no contribution other than E1, 
the E2 and M1 transitions are no longer insignificant at $E=0.5$ MeV.

\begin{figure}[h]
\centerline{
	\parbox[t]{.45\textwidth}{
	\psfig{figure=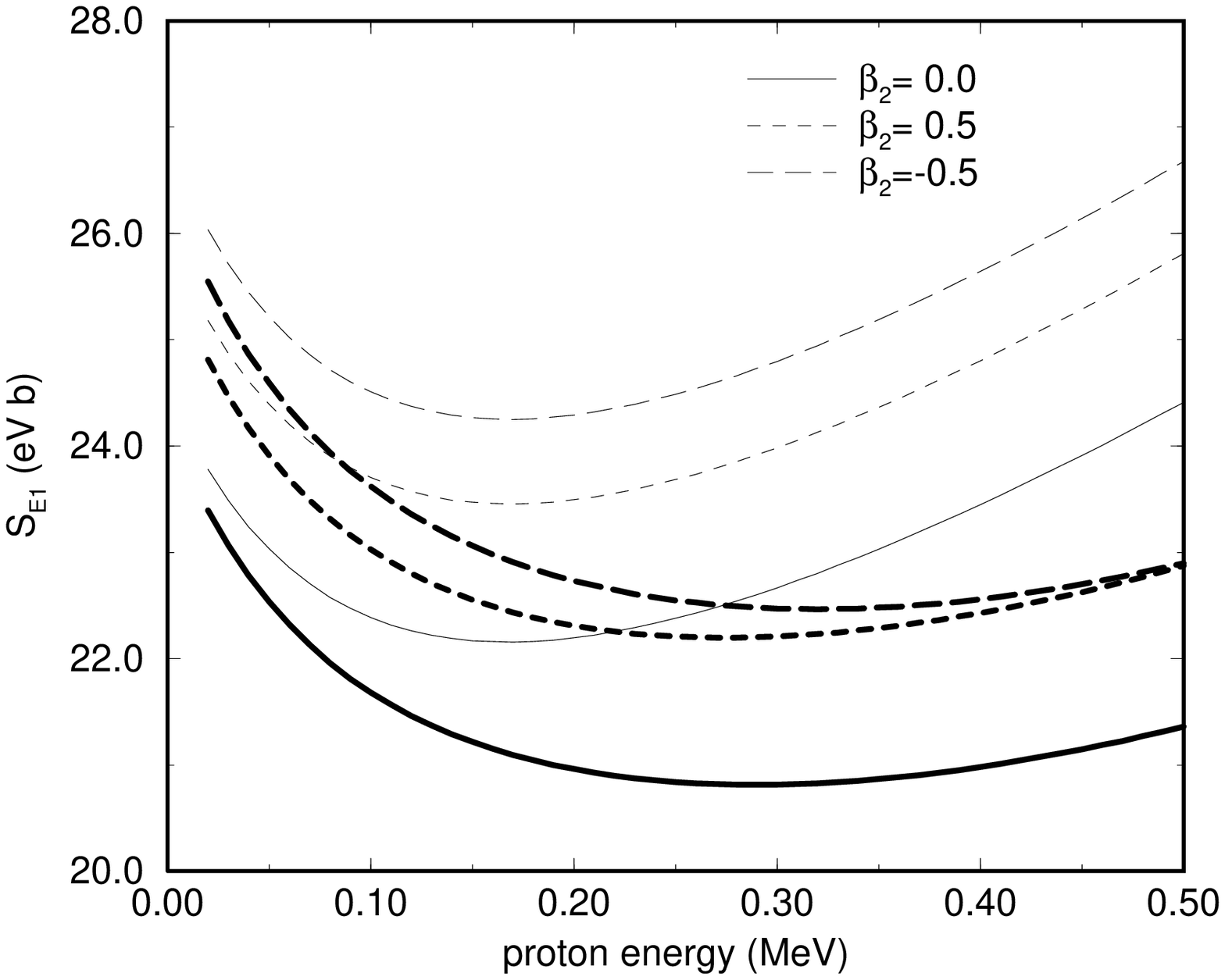,width=0.45\textwidth}
	\caption{The effect of nuclear phase shifts on the low energy S-factor
when considering only reorientation couplings for the g.s. of $^7$Be:
thick lines include the nuclear interaction for the scattering states.}	
	\label{b8ae1fsi}}
\hspace{0.05\textwidth}
	\parbox[t]{.45\textwidth}{
	\psfig{figure=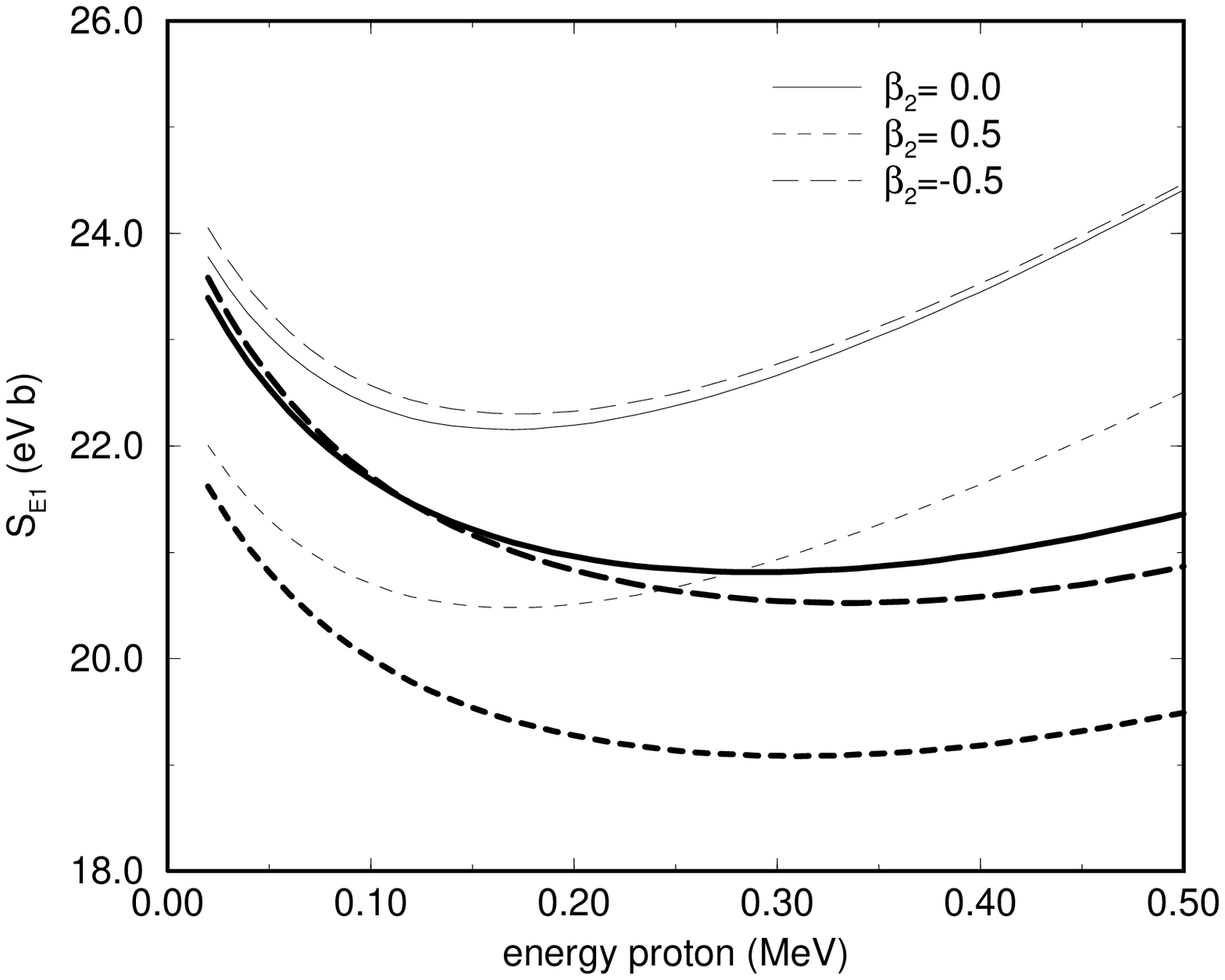,width=0.45\textwidth}
	\caption{The effect of nuclear phase shifts on the low energy S-factor
when considering couplings to excited states of $^7$Be:
thick lines include the nuclear interaction for the scattering states.}	
	\label{b8ce1fsi}}
}
\end{figure}

\begin{figure}[h]
\centerline{
	\parbox[t]{0.7\textwidth}{
\centerline{\psfig{figure=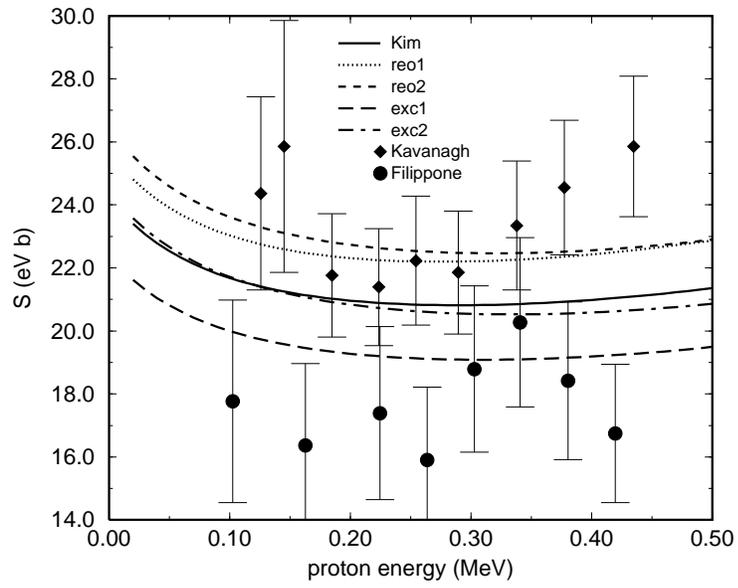,width=0.7\textwidth}}
	\caption{The comparison of $S_{E1}(E)$ including
the nuclear interaction in the scattering states with the low energy data.}	
\label{b8se1data}}
}
\end{figure}

In order to estimate the significance of
nuclear interactions in the continuum, we performed calculations assuming that the
negative parity continuum states would be subject to the same p-core potential used to obtain the g.s. wavefunction given that nothing is known about their nuclear phase shifts.
In table \ref{e1fsitb} we present $S_{E1}$ for $E=20$ keV and $E=500$ keV,
followed by the ratio $R=\frac{S(20)}{S(500)}$ for Kim's model, the reorientation models
(both $\beta_2=0.5$ and $\beta_2=-0.5$), and the excitation models
(both $\beta_2=0.5$ and $\beta_2=-0.5$). The rows  with * correspond
to calculations including nuclear phase shifts in the scattering states
(same interaction as the g.s. interaction).
The nuclear interaction in the continuum can produce up to $15\%$ effect
on the low energy behaviour of $S_{E1}$. This corresponds to an upper 
limit for the uncertainty induced by the lack of information for the structure
of the negative parity states.
In fig.(\ref{b8ae1fsi}) we illustrate the modification in
the low energy behaviour of $S_{E1}$ caused by the nuclear interaction in the continuum for
the models that only include the g.s. of the core.
In fig.(\ref{b8ce1fsi}) we present the equivalent for the models with core
excitation. We point out that when including nuclear phase shifts in
the scattering states, there is some sensitivity to the core's shape and structure.
This is illustrated in fig.(\ref{b8se1data}) where
we compare the low energy $S_{E1}(E)$ plots for the set of
models with the experimental results: 
Kim's original model, the reorientation models and the core excitation
models. The results shown are from calculations which 
 include nuclear phase shifts but only E1 transitions. 
We include
in fig.(\ref{b8se1data}) two sets of data available in
this energy region \cite{filippone,kavanagh} normalised
to the most recent $^7$Li(d,p) cross section \cite{strieder}.

In conclusion we 
have extended calculations for the g.s. of $^8$B based on the $core+p$ model
where the core is allowed to deform and excite. 
The density distribution and the momentum distribution 
are not very sensitive to the range of models.
The quadrupole and the magnetic moments are strongly dependent on the
core's structure but in order to make predictions for $^8$B it is
important to measure the quadrupole and magnetic moment of the core.
We found that the overall normalisation of the S-factor is modified 
with the core's structure, as well as with  the shape of the Woods-Saxon g.s. interaction.
In all cases the {\em shape} of the energy behaviour is not significantly changed.
Finally we maintain that the uncertainties associated with the $^7$Be$-p$
phase shifts(the negative parity states)
introduce an uncertainty ($< 15\%$) into the S-factor energy behaviour.

\vspace{0.5cm}
\noindent
This work was supported by the JNICT grant PBIC/C/FIS/2155/95 and BIC 1481.

\newpage

\begin{table}[h]
\centering
 \begin{tabular}{||l|r|r|r|r|r|r|r||} \hline   
 Model &  $\beta_2$  & $\beta_3$  & Core States & $V_{ws}$ (MeV)  & 
$P[p_{3/2}\otimes\frac{3}{2}] $ & $P[I=\frac{1}{2}^-] $ & $P[I=\frac{7}{2}^-] $
\\ \hline\hline
Kim \cite{kim} &        $0.0$ & $0.0$  & $\frac{3}{2}^-$ & $-31.768$ &  
	$1.00$ & $0.00$ & $0.00$   
\\ \hline
reo1 &        $0.5$ & $0.0$  & $\frac{3}{2}^-$ & $-32.5667$ &  
	$0.57$ & $0.00$ & $0.00$    
\\ \hline
reo2 &        $-0.5$ & $0.0$  & $\frac{3}{2}^-$ & $-29.2343$ &  
	$0.94$ & $0.00$ & $0.00$    
\\ \hline
exc1 &        $0.5$ & $0.0$  & $\frac{3}{2}^-;\frac{1}{2}^-;\frac{7}{2}^-$ & $-29.9156$ &  
	$0.39$ & $0.03$ & $0.14$    
\\ \hline
exc2 &        $-0.5$ & $0.0$  & $\frac{3}{2}^-;\frac{1}{2}^-;\frac{7}{2}^-$ & $-28.5198$ &  
	$0.87$ & $0.09$ & $0.01$    
\\ \hline
oct1 &        $0.5$ & $0.5$  & $\frac{3}{2}^-;\frac{1}{2}^-;\frac{7}{2}^-$ & $-30.1736$ &  
	$0.41$ & $0.04$ & $0.13$    
\\ \hline
oct2 &        $-0.5$ & $0.5$  & $\frac{3}{2}^-;\frac{1}{2}^-;\frac{7}{2}^-$ & $-28.5003$ &  
	$0.87$ & $0.09$ & $0.01$    
\\ \hline \hline
\end{tabular}
\caption{Deformed core models for the g.s. of $^8$B: parameters for the nuclear interaction depth  (for the other parameters we use Kim's model:
$V_{so}=-2.06$, $R=2.95$ fm and $a=0.52$ fm), the core's structure and the probabilities 
for the main channels ($I$ is the spin of the core). 
}
\label{pardeftb}
\end{table}

\begin{table}[h]
\centering
 \begin{tabular}{||l|r|r||} \hline   
 Model &  $\sqrt{\langle r_{valence}^2} \rangle$ & $\Gamma$  \\ 
    &     (fm) & (MeV/c) \\ \hline\hline
Kim \cite{kim} &         $4.623$ & $158$   
\\ \hline
reo1 &        $4.72$ & $154$   
\\ \hline
reo2 &        $4.78$ & $154$   
\\ \hline
exc1 &        $4.64$ & $158$   
\\ \hline
exc2 &       $4.75$ & $154$   
\\ \hline
oct1 &        $4.76$ & $154$   
\\ \hline
oct2 &        $4.81$ & $150$   
\\ \hline \hline
\end{tabular}
\caption{Deformed core models for the g.s. of $^8$B: 
the valence proton contribution to the predicted r.m.s. matter radius and 
the momentum distribution width. 
}
\label{deftb}
\end{table}

\begin{table}[h]
\centering
 \begin{tabular}{||l|r|r||} \hline   
 Model &  $ Q_{R} $ & $\mu_R$  \\ 
    &     (e fm) & ($\mu_B$) \\ \hline\hline
Kim \cite{kim} &         $0$ & $2.53$   
\\ \hline
reo1 &        $-7.45$ & $-0.18$   
\\ \hline
reo2 &        $3.78$ & $3.09$   
\\ \hline
exc1 &        $-7.07$ & $-0.62$   
\\ \hline
exc2 &       $2.26$ & $3.08$   
\\ \hline
oct1 &        $-7.52$ & $-0.50$   
\\ \hline
oct2 &        $2.30$ & $3.07$   
\\ \hline \hline
\end{tabular}
\caption{Quadrupole and Magnetic moments for the g.s. of $^8$B: 
contributions from the valence proton. 
}
\label{qmtb}
\end{table}

\begin{table}[h]
\centering
 \begin{tabular}{||l|r|r|r||} \hline   
 Model &    S(100 keV) & S(20 keV) & $\frac{S(20)}{S(100)}$\\
   &    (eV b)     & (eV b) &   \\ \hline\hline
Kim &          $22.4$ & $23.8$ & $1.06$  \\ \hline
reo1 &   $23.7$ & $25.2$ & $1.06$  \\ \hline
reo2 &    $24.5$ &$26.0$ & $1.06$  \\ \hline
exc1 &    $20.7$ &$22.0$ & $1.06$  \\ \hline
exc2 &       $22.6$ &$24.1$ & $1.06$   \\ \hline \hline
\end{tabular}
\caption{The S-factors at $20$ keV and $100$ keV with no
nuclear phase shifts (using $R_{max}=360$ fm) and the S-factor ratio.}
\label{e1tb}
\end{table}

\begin{table}[h]
\centering
 \begin{tabular}{||l|r|r|r|r||} \hline   
 Model &    S(500 keV) & S(20 keV) & $R=\frac{S(20)}{S(500)}$ & $\frac{\Delta R}{R}$ \\
   &    (eV b)     & (eV b) &   & \\ \hline\hline
Kim   &   $24.4$ & $23.8$ & $0.97$ &  \\ \hline
Kim*   &   $21.4$ & $23.4$ & $1.10$ &  $12.4$ \% \\ \hline
reo1 &   $25.8$ & $25.2$ & $0.98$ &  \\ \hline
reo1*   &   $22.9$ & $24.8$ & $1.09$ &  $11.2$ \% \\ \hline
reo2 &   $26.7$ & $26.0$ & $0.98$ &  \\ \hline
reo2*   &   $22.9$ & $25.5$ & $1.12$ &  $14.3$ \% \\ \hline
exc1 &   $22.5$ & $22.0$ & $0.98$ &  \\ \hline
exc1*   &   $19.5$ & $21.6$ & $1.11$ &  $13.4$ \% \\ \hline
exc2 &   $24.5$ & $24.0$ & $0.98$ &  \\ \hline
exc2*   &   $20.9$ & $23.6$ & $1.13$ &  $15.0$ \% \\ \hline \hline
\end{tabular}
\caption{The influence of nuclear phase shifts in the
S-factors at $20$ keV and $500$ keV and the S-factor low-energy behaviour
(* including the nuclear phase shifts in the scattering states).}
\label{e1fsitb}
\end{table}

\end{document}